\documentclass[12pt]{article}

\begin{document}


\def\a{\alpha}
\def\b{\beta}
\def\c{\varepsilon}
\def\d{\delta}
\def\e{\epsilon}
\def\f{\phi}
\def\g{\gamma}
\def\h{\theta}
\def\k{\kappa}
\def\l{\lambda}
\def\m{\mu}
\def\n{\nu}
\def\p{\psi}
\def\q{\partial}
\def\r{\rho}
\def\s{\sigma}
\def\t{\tau}
\def\u{\upsilon}
\def\v{\varphi}
\def\w{\omega}
\def\x{\xi}
\def\y{\eta}
\def\z{\zeta}
\def\D{\Delta}
\def\G{\Gamma}
\def\H{\Theta}
\def\L{\Lambda}
\def\F{\Phi}
\def\P{\Psi}
\def\S{\Sigma}
\def\V{\varPsi}

\def\o{\over}
\newcommand{\sla}[1]{#1 \llap{\, /}}

\newcommand{\beq}{\begin{eqnarray}}
\newcommand{\eeq}{\end{eqnarray}}
\newcommand{\gsim}{ \mathop{}_{\textstyle \sim}^{\textstyle >} }
\newcommand{\lsim}{ \mathop{}_{\textstyle \sim}^{\textstyle <} }
\newcommand{\vev}[1]{ \left\langle {#1} \right\rangle }
\newcommand{\bra}[1]{ \langle {#1} | }
\newcommand{\ket}[1]{ | {#1} \rangle }

\newcommand{\EV}{ {\rm eV} }
\newcommand{\KEV}{ {\rm keV} }
\newcommand{\MEV}{ {\rm MeV} }
\newcommand{\GEV}{ {\rm GeV} }
\newcommand{\TEV}{ {\rm TeV} }


\baselineskip 0.7cm

\begin{titlepage}

\begin{flushright}
UT-03-16 \\
\end{flushright}

\vskip 1.35cm
\begin{center}
{\large \bf
Supergravity Minimal Inflation \\
and its Spectral Index Revisited
}
\vskip 1.2cm
Izawa K.-I.
\vskip 0.4cm

{\it Department of Physics, University of Tokyo,\\
     Tokyo 113-0033, Japan}

\vskip 1.5cm

\abstract{
Natural supergravity models of new
inflation are reconsidered
as minimal inflationary models
within slow-roll approximation.
Their running spectral index is derived
in a revised form
with recent observational results
and future refinements in mind.
This will possibly determine essential model
parameters with respect to Planck-suppressed operators.
}
\end{center}
\end{titlepage}

\setcounter{page}{2}


Primordial inflation
\cite{Lyt}
is expected to open a window into
Planck-suppressed physics:
Among possible ingredients beyond the standard model,
supersymmetry roughly achieves natural
realization of a flat potential for an inflaton.
Discrepancy from the complete flatness
may reflect physics suppressed by the gravitational scale
\cite{Iza}.
In this respect, the detections of primordial fluctuations
of cosmic microwave background radiation
initiated by COBE and continued by WMAP, to name a few
\cite{Ben},%
\footnote{See also recent investigations on inflationary models
in Refs.\cite{Fen, Kaw}.}
might provide evidence for such basic objects
as effective operators of higher dimensions with Planck-scale cutoff,
which are naturally expected in the universal framework
of effective field theories. 

As a simplest example, we consider a single-field
model for slow-roll inflation with a potential
\begin{equation}
 V(\v) = v^4 - {\k \o 2}v^4\v^2
              - {\l \o n!}\v^n
 \label{POT}
\end{equation}
for $0 < \l, v, \k, \v \ll 1$ and $n \geq 3$.
Here and henceforth, we adopt the unit with the reduced Planck scale
equal to one.
The first term $v^4$ yields vacuum energy for inflation;
the second one ${\k \o 2}v^4\v^2$ affects slow-roll dynamics
during inflation; and the last one ${\l \o n!}\v^n$
eventually terminates inflation.
Note that the initial condition for the inflaton field $\v$
may be set by primary inflation
\cite{Lyt, Yan, Ken}.

This form is ubiquitous in supergravity inflation.
For instance, 
let us take a minimal model considered in
Ref.\cite{Iza}.
By means of a single chiral superfield $\f$,
an inflaton $\v$ can be provided by
$\sqrt{2}$ times the real part of its lowest component.
We adopt a natural superpotential%
\footnote{This form is protected by nonrenormalization
or $R$ symmetry.}
\begin{equation}
 W = v^2 \f - {g \o n+1} \f^{n+1},
 \label{ADDEQ}
\end{equation}
where $g > 0$, and a generic K\"ahler potential
\begin{equation}
 K = |\f|^2 + {\k \o 4}|\f|^4 + \cdots,
\end{equation}
where
$\k$ is generated, at least, radiatively of order $10^{-2}$
due to suppression by a loop factor
and the ellipsis denotes higher-order terms
which may be disregarded.%
\footnote{Concrete forms of radiative
corrections lie beyond the scope of this paper.}
The tiny scale $v^2$ can be generated dynamically
\cite{Iza, Hot}.

The potential for the lowest component $\f$
is given in supergravity by
\begin{equation}
  V = e^K
      \left\{ \left( {\q^2 K \o \q \f \q \f^*} \right)^{-1} |DW|^2
      - 3|W|^2 \right\},
 \label{EPOT}
\end{equation}
where we have defined
\begin{equation}
 DW = {\q W \o \q \f} + {\q K \o \q \f}W.
\end{equation}
Thus the potential is approximately given by Eq.(\ref{POT}) with
\cite{Iza}
\beq
 {\l \o n!}={g \o 2^{{n \o 2}-1}}v^2.
\eeq
The above example clearly shows that the $\v$-dependent terms in
Eq.(\ref{POT}) directly originate from Planck-suppressed interactions
with operators of higher dimensions.

This type of potentials also appears
at the second stage of the double inflation
\cite{Kaw, Yan, Sug}.
Furthermore, $R$-invariant models may also yield such potentials effectively
\cite{Asa}.

Let us investigate the inflationary dynamics of the above model
Eq.(\ref{POT}) within slow-roll approximation
\cite{Lyt}.

The slow-roll inflationary regime is determined by the condition
\begin{equation}
 \e(\v) = {1 \o 2} \left({V'(\v) \o V(\v)} \right)^2 \leq 1,
 \quad |\y(\v)| \leq 1,
 \label{COND}
\end{equation}
where
\begin{equation}
 \y(\v) = {V''(\v) \o V(\v)}.
\end{equation}
For the potential Eq.(\ref{POT}), we obtain
\begin{equation}
\begin{array}{l}
 \displaystyle
 \e(\v) \simeq {1 \o 2}
          \left({-\k v^4\v - {\l \o {(n-1)!}}\v^{n-1} \o v^4}
            \right)^2
        = {\v^2 \o 2}\left(\k + {\l v^{-4} \o {(n-1)!}}\v^{n-2}\right)^2, \\
 \noalign{\vskip 2ex}
 \displaystyle
 \y(\v) \simeq {-\k v^4 - {\l \o {(n-2)!}}\v^{n-2} \o v^4}
        = -\k - {\l v^{-4} \o {(n-2)!}}\v^{n-2}.
\end{array}
\end{equation}
The slow-roll condition Eq.(\ref{COND}) is satisfied
for $\v \leq \v_f$
where
\begin{equation}
 \v_f^{n-2} \simeq {(n-2)!(1-\k) \o \l v^{-4}},
\end{equation}
which provides the value of the inflaton field at the end of inflation.

The value $\v_N$ of the inflaton field
corresponding to the $e$-fold number $N$ is given by
\beq
 N &=& \int_{\v_f}^{\v_N} \! d\v \, {V(\v) \o V'(\v)}
 \nonumber \\
 &\simeq& \int_{\v_f}^{\v_N} \! d\v \,
 {v^4 \o -\k v^4\v - {\l \o {(n-1)!}}\v^{n-1}},
\eeq
which implies
\beq
 \v_N^{n-2} \simeq {(n-1)!\k \o \l v^{-4}}\left(
 {1+(n-2)\k \o 1-\k}e^{(n-2)\k N} - 1 \right)^{-1}.
\eeq
Hence the scalar spectral index $n_s$ of the primordial density fluctuations
\cite{Lyt}
is given by
\beq
  n_s(N) &\simeq& 1 - 6\e(\v_N) + 2\y(\v_N) \nonumber \\
  &\simeq& 1 - 2\k\left\{1+(n-1)\left(
  {1+(n-2)\k \o 1-\k}e^{(n-2)\k N} - 1 \right)^{-1}
  \right\}.
 \label{NSN}
\eeq
This expression is intended
for comparison with anticipated precise data.
Note that it is independent of the coupling $\l$ and the scale $v$,
and crudely $n_s \simeq 1-2\k$.
Sizable deviations of the value $n_s$ from one
thus directly indicate the presence of
Planck-suppressed effects stemming from higher-dimensional operators
in the K{\" a}hler potential, in particular.

The amount of running with respect to the $e$-fold number $N$
or the logarithm of the comoving wavenumber%
\footnote{Numerical correspondence between these two variables
depends on the reheating process, which we do not specify
in this paper.}
can be seen immediately from Eq.(\ref{NSN}).
For the sake of convenience,
numerical values of the running spectral index $n_s(N)$
for the cases $n=3,4,5$ are given in Tables 1,2,3, respectively.

That is, in the realm of precision cosmology,
we might obtain $\k$, $n$, and so on with some fundamental meaning
from analyses of observational results,
if the class Eq.(\ref{POT}), for instance, is realized in Nature.
The values $n_s(N)$ in the Tables are largely consistent with
the recent WMAP data,
though the class Eq.(\ref{POT}) may well turn out to be,
at least, incomplete in no distant future,
since its predictions are quite restricted due to its minimality.
Then, more elaborate models with
two or more fields in collaboration
\cite{Lyt, Fen, Kaw, Yan, Sug}
might be adequate.
Even in such a case, it is conceivable that
some effects suppressed by the gravitational scale will be revealed
through detailed analyses of observational data
in the near future. This, if true, confirms our view that
the framework of effective field theories is universal.

\section*{Acknowledgements}

We would like to thank T.~Watari and T.~Yanagida for valuable discussions.


\newpage

\begin{table}

\begin{center}

\begin{tabular}{|r||r|r|r|r|r|r|r|}
\hline
$e$-fold $N$ & 60 & 55 & 50 & 45 & 40 & 35 & 30 \\
\hline
\hline
$\k=.01$ & .933 & .928 & .921 & .913 & .903 & .891 & .874 \\
\hline
.02 & .927 & .922 & .916 & .909 & .899 & .887 & .871 \\
\hline
.03 & .918 & .914 & .908 & .901 & .892 & .881 & .866 \\
\hline
.04 & .905 & .902 & .897 & .891 & .883 & .873 & .858 \\
\hline
.05 & .891 & .888 & .884 & .879 & .872 & .863 & .849 \\
\hline
.06 & .874 & .872 & .869 & .865 & .859 & .851 & .839 \\
\hline
.07 & .856 & .855 & .852 & .849 & .844 & .837 & .827 \\
\hline
\end{tabular}

\end{center}

\caption{The running spectral index $n_s$ for $n=3$.}

\end{table}

\begin{table}

\begin{center}

\begin{tabular}{|r||r|r|r|r|r|r|r|}
\hline
$e$-fold $N$ & 60 & 55 & 50 & 45 & 40 & 35 & 30 \\
\hline
\hline
$\k=.01$ & .955 & .951 & .947 & .941 & .934 & .924 & .912 \\
\hline
.02 & .949 & .946 & .942 & .938 & .932 & .924 & .912 \\
\hline
.03 & .935 & .934 & .931 & .928 & .924 & .917 & .908 \\
\hline
.04 & .918 & .917 & .916 & .914 & .911 & .906 & .899 \\
\hline
.05 & .899 & .899 & .898 & .897 & .895 & .892 & .887 \\
\hline
.06 & .880 & .880 & .879 & .879 & .877 & .875 & .872 \\
\hline
.07 & .860 & .860 & .860 & .859 & .859 & .857 & .855 \\
\hline
\end{tabular}

\end{center}

\caption{The running spectral index $n_s$ for $n=4$.}

\end{table}

\begin{table}

\begin{center}

\begin{tabular}{|r||r|r|r|r|r|r|r|}
\hline
$e$-fold $N$ & 60 & 55 & 50 & 45 & 40 & 35 & 30 \\
\hline
\hline
$\k=.01$ & .965 & .962 & .958 & .953 & .947 & .939 & .929 \\
\hline
.02 & .956 & .954 & .952 & .949 & .945 & .940 & .931 \\
\hline
.03 & .939 & .938 & .938 & .936 & .934 & .930 & .925 \\
\hline
.04 & .920 & .920 & .919 & .919 & .918 & .916 & .912 \\
\hline
.05 & .900 & .900 & .900 & .900 & .899 & .898 & .896 \\
\hline
.06 & .880 & .880 & .880 & .880 & .880 & .879 & .878 \\
\hline
.07 & .860 & .860 & .860 & .860 & .860 & .860 & .859 \\
\hline
\end{tabular}

\end{center}

\caption{The running spectral index $n_s$ for $n=5$.}

\end{table}


\begin{thebibliography}{99}

\bibitem{Lyt} For reviews, D.H.~Lyth and A.~Riotto, arXiv:hep-ph/9807278; \\
              W.H.~Kinney, arXiv:astro-ph/0301448; \\
              S.F.~King, arXiv:hep-ph/0304264.

\bibitem{Iza} Izawa K.-I.~and T.~Yanagida, arXiv:hep-ph/9608359.

\bibitem{Ben} C.L.~Bennett {\it et al.}, arXiv:astro-ph/9601067; \\
              S.~Hannestad, S.H.~Hansen, and F.L.~Villante,
              arXiv:astro-ph/0012009; \\
              C.L.~Bennett {\it et al.}, arXiv:astro-ph/0302207; \\
              D.N.~Spergel {\it et al.}, arXiv:astro-ph/0302209; \\
              E.~Komatsu {\it et al.}, arXiv:astro-ph/0302223; \\
              H.V.~Peiris {\it et al.}, arXiv:astro-ph/0302225; \\
              S.L.~Bridle, A.M.~Lewis, J.~Weller, and G.~Efstathiou,
              arXiv:astro-ph/0302306; \\
              V.~Barger, H.-S.~Lee, and D.~Marfatia, arXiv:hep-ph/0302150; \\
              C.R.~Contaldi, H.~Hoekstra, and A.~Lewis,
              arXiv:astro-ph/0302435; \\
              U.~Seljak, P.~McDonald, and A.~Makarov,
              arXiv:astro-ph/0302571; \\
              P.~Mukherjee and Y.~Wang, arXiv:astro-ph/0303211; \\
              W.A.~Chiu, X.~Fan, and J.P.~Ostriker, arXiv:astro-ph/0304234.

\bibitem{Fen} B.~Feng, M.~Li, R.-J.~Zhang, and X.~Zhang,
              arXiv:astro-ph/0302479; \\
              B.~Kyae and Q.~Shafi, arXiv:astro-ph/0302504; \\
              J.~Ellis, M.~Raidal, and T.~Yanagida, arXiv:hep-ph/0303242; \\
              L.~Pogosian, S.-H.H.~Tye, I.~Wasserman, and M.~Wyman,
              arXiv:hep-th/0304188; \\
              Q.-G.~Huang and M.~Li, arXiv:hep-th/0304203; \\
              W.H.~Kinney, E.W.~Kolb, A.~Melchiorri, and A.~Riotto,
              arXiv:hep-ph/0305130.

\bibitem{Kaw} M.~Kawasaki, M.~Yamaguchi, and J.~Yokoyama,
              arXiv:hep-ph/0304161.

\bibitem{Yan} Izawa K.-I., M.~Kawasaki, and T.~Yanagida,
              arXiv:hep-ph/9707201.

\bibitem{Ken} Izawa K.-I., arXiv:hep-ph/9710479.

\bibitem{Hot} Izawa~K.-I. and T.~Yanagida, arXiv:hep-th/9602180;
              arXiv:hep-ph/9809366; \\
              arXiv:hep-ph/9904426; \\
              T.~Hotta, Izawa~K.-I., and T.~Yanagida,
              arXiv:hep-ph/9606203; \\
              Izawa K.-I., Y.~Nomura, K.~Tobe, and T.~Yanagida,
              arXiv:hep-ph/9705228; \\
              Izawa K.-I., arXiv:hep-ph/9708315.

\bibitem{Sug} M.~Kawasaki, N.~Sugiyama, and T.~Yanagida,
              arXiv:hep-ph/9710259; \\
              M.~Kawasaki and T.~Yanagida, arXiv:hep-ph/9807544; \\
              T.~Kanazawa, M.~Kawasaki, N.~Sugiyama, and T.~Yanagida,
              arXiv:hep-ph/9908350; arXiv:astro-ph/0006445; \\
              T.~Kanazawa, M.~Kawasaki, and T.~Yanagida,
              arXiv:hep-ph/0002236.

\bibitem{Asa} Izawa K.-I., M.~Kawasaki, and T.~Yanagida,
              arXiv:hep-ph/9810537; \\
              T.~Asaka, K.~Hamaguchi, M.~Kawasaki, and T.~Yanagida,
              arXiv:hep-ph/9906366; arXiv:hep-ph/9907559; \\
              M.~Kawasaki, N.~Sakai, M.~Yamaguchi, and T.~Yanagida,
              arXiv:hep-ph/0005073.

\end{thebibliography}
\end{document}